\documentclass[conference]{IEEEtran}
\IEEEoverridecommandlockouts
\usepackage{cite}
\usepackage{amsmath,amssymb,amsfonts}

\usepackage{graphicx}
\usepackage{textcomp}
\usepackage{xcolor}
\usepackage{color}

\usepackage{algpseudocode}
\usepackage{algorithm}

\algnewcommand\algorithmicforeach{\textbf{for each}}
\algdef{S}[FOR]{ForEach}[1]{\algorithmicforeach\ #1\ \algorithmicdo}

\def\BibTeX{{\rm B\kern-.05em{\sc i\kern-.025em b}\kern-.08em
    T\kern-.1667em\lower.7ex\hbox{E}\kern-.125emX}}

\IEEEpubid{\begin{minipage}{\textwidth}
        \centering\footnotesize{\vspace{3.5cm} © 2023 IEEE. Personal use of this material is permitted. Permission from IEEE must be
obtained for all other uses, in any current or future media, including
reprinting/republishing this material for advertising or promotional purposes, creating new collective works, for resale or redistribution to servers or lists, or reuse of any copyrighted
component of this work in other works.}
\end{minipage}}     
\begin{document}

\title{Beam Management Driven by Radio Environment Maps in O-RAN Architecture
\thanks{The presented work has been funded by the Polish Ministry of Education and Science within the status activity task no. 0312/SBAD/8164.}
}

\author{\IEEEauthorblockN{Marcin Hoffmann}
\IEEEauthorblockA{\textit{Rimedo Labs}}
\IEEEauthorblockA{\textit{Institute of Radiocommunications} \\
\textit{Poznan University of Technology}\\
Poznan, Poland \\
0000-0002-4957-0173}
\and
\IEEEauthorblockN{Pawel Kryszkiewicz}
\IEEEauthorblockA{\textit{Rimedo Labs}}
\IEEEauthorblockA{\textit{Institute of Radiocommunications} \\
\textit{Poznan University of Technology}\\
Poznan, Poland \\
0000-0001-9054-9416}
}

\maketitle

\begin{abstract}
The Massive Multiple-Input Multiple-Output (M-MIMO) is considered as one of the key technologies in 5G, and future 6G networks. From the perspective of, e.g., channel estimation, especially for high-speed users it is easier to implement an M-MIMO network exploiting a static set of beams, i.e., Grid of Beams (GoB). While considering GoB it is important to properly assign users to the beams, i.e., to perform Beam Management (BM). BM can be enhanced by taking into account historical knowledge about the radio environment, e.g., to avoid radio link failures.
The aim of this paper is to propose such a BM algorithm, that utilizes location-dependent data stored in a Radio Environment Map (REM). 
 It utilizes received power maps, and user mobility patterns to optimize the BM process in terms of Reinforcement Learning (RL) by using the Policy Iteration method under different goal functions, e.g., maximization of received power or minimization of beam reselections while avoiding radio link failures.
 The proposed solution is compliant with the Open Radio Access Network (O-RAN) architecture, enabling its practical implementation.
Simulation studies have shown that the proposed BM algorithm can significantly reduce the number of beam reselections or radio link failures 
compared to the baseline algorithm. 

\end{abstract}


\section{Introduction}

Massive Multiple-Input Multiple-Output (M-MIMO) is considered one of the key enablers for achieving high user throughput in contemporary 5G, and future 6G networks~\cite{akyildiz2020}. The idea of M-MIMO is to equip Base Stations (BSs) with large antenna arrays, i.e., exploiting dozens or even hundreds of antennas. Through proper amplitude scaling and phase shifting of the signal being transmitted or received by each of BS's antennas the radiation pattern of the antenna array can be modified in a flexible manner~\cite{Ehab2017}. This procedure is known as beamforming. Beamforming allows for, e.g., increasing received signal strength, or suppressing interference. Moreover, due to the application of beamforming, it is possible to simultaneously serve multiple User Equipments (UEs) using the same time-frequency resources, i.e., spatial domain multiple access. There are two main approaches to the practical implementation of the M-MIMO system, namely Grid of Beams (GoB) beamforming, and adaptive beamforming~\cite{Lee2021}. GoB beamforming in 5G relies on downlink measurements of the Reference Signal Received Power (RSRP), based typically on Synchronization Signal Blocks (SSBs) transmitted by BS \cite{Lee2021}. On the opposite, adaptive beamforming requires accurate channel estimation, which in a time-division duplexing (TDD) system is based on the Sounding Reference Signal (SRS) transmitted by UE in the uplink. Although SRS-based beamforming allows achieving higher user throughputs, due to better knowledge about radio channel coefficients, it requires more complex signal processing and management of SRS signals transmitted by UEs in the neighboring cells. Moreover, SRS-based beamforming is inappropriate for high-speed UEs because the channel estimate becomes almost immediately outdated when UE moves~\cite{Davydov2021}. From this perspective, GoB beamforming that utilizes a static set of beams, requiring UE to only indicate RSRP related to SSB, is much easier for practical implementations. GoB beamforming is indicated as a promising solution to be implemented in millimeter wave frequency bands~\cite{Rebato2019}.  

GoB beamforming exploits a static set of beams covering the area of a cell. However, due to reflections, or shadowing, e.g., from surrounding buildings, RSRP related to a particular beam can follow an unobvious spatial distribution. Moreover, while considering high-speed UEs the RSRP can change rapidly, requiring fast beam reselections supported by the 5G control plane. Following the specifications of the 5G, BSs send SSB signals for all the covered beams in a 5~ms long burst. The burst repeats every 20~ms~\cite{barati2020energy}. The most straightforward approach to BM would be to select a new beam after every SSB burst. However, while switching between beams within a single cell requires less Radio Resource Control (RRC) configuration than the inter-cell handover the signaling, and reconfiguration overhead still exists~\cite{jo2020}. Some balance between beam reselections and signaling overhead should be found.  From this perspective, Beam Management (BM) is indicated as one of the key challenges for GoB beamforming~\cite{Li2020}. 

In \cite{Abinader2021}, a baseline, 5G BM algorithm is proposed. Therein, the beam is switched when RSRP related to the best target beam is above the current source beam by some specified margin. Such an approach can prevent too frequent beam reselections, but it is characterized by poor flexibility by a single margin parameter. To overcome these issues, especially in the context of high-speed UEs, some studies propose to utilize context information to support BM, e.g., UE location, speed, and bearing. The authors of~\cite{Na2019} propose to train a so-called Deep Learning Agent that takes UE location, Signal-to-Noise-Ratio (SNR), mobility vectors, and current beam index, to predict if RSRP would fall below the desired level. Based on that prediction, the decision on beam switching is made. The drawback of the method is that the decision is greedy, i.e., no long-term optimization taking into account UE's expected route is applied. In~\cite{Kalamkar2022}, the authors proposed a stochastic geometry modeling of BM to maximize the area spectral efficiency. However, to formulate and resolve the optimization problem, some simplifications are assumed, e.g., with regard to the antenna array radiation pattern, BSs placement, and channel model. These can make the proposed solution impractical in real environments. Finally, there are some algorithms focused on the handovers in 4G-LTE networks utilizing the so-called Radio Environment Maps (REMs), e.g.,~\cite{Suarez2017}. The proposed REM stores RSRP-location pairs in the form of a map, to be used to predict the signal level for UE in the next time step, and decide on the eventual handover. However, the algorithm is designed to deal with handover between two BSs equipped with omnidirectional antennas. It is not adequate for BM in a 5G network utilizing GoB beamforming. In such a case, UE can be dynamically switched between up to tens of candidate beams. 
Last but not least, most authors of the BM algorithms do not consider what signals and Key Performance Indicators (KPIs) have to be exchanged in 5G networks. Moreover, there is no discussion on how their proposal can be implemented in a real network. The lack of such considerations can be justified by the fact that contemporary large-scale mobile networks are not appropriate for introducing novel, third-party ideas. There are limited possibilities for controlling Radio Access Network (RAN), e.g., by sending messages related to BM or receiving RSRP measurement reports. However, in recent years the idea of the so-called Open-Radio Access Network (O-RAN) is gaining more and more attention~\cite{Bonati2021}.

In this paper, we propose a REM-based BM adjusted for the O-RAN architecture. REM stores two types of location-dependent data: RSRP related to each beam at a given location, and a UE mobility map that describes the motion pattern of UEs in a probabilistic manner. Based on these two types of maps we define BM as a Markov Decision Process (MDP) that can be optimized with the use of the so-called Policy Iteration (PI) method from the family of RL algorithms, aimed at the maximization of a cumulative reward (related to the optimization goal)~\cite{sutton2018reinforcement}. The process is flexible and allows the Mobile Network Operator (MNO) to define its optimization goals, e.g., maximization of RSRP, and minimization of beam reselections, while maintaining the desired RSRP. 
Most importantly, our solution is compliant with the O-RAN architecture enabling its practical implementation. 

The paper is organized as follows: The system model is described in Sec.~\ref{sec:system_model}. The required components placement in the O-RAN architecture is presented in Sec.~\ref{sec:oran_architecture}. The proposed REM-based BM algorithm is presented in Sec.~\ref{sec:rem-bm}. The simulation scenario is provided in Sec.~\ref{sec:scenario} with results discussed in Sec.~\ref{sec:results}. The paper is concluded in Sec.~\ref{sec:conclusions}. 

\section{System Model} \label{sec:system_model}
\begin{figure}[htbp]
\centerline{\includegraphics[height=4.2cm]{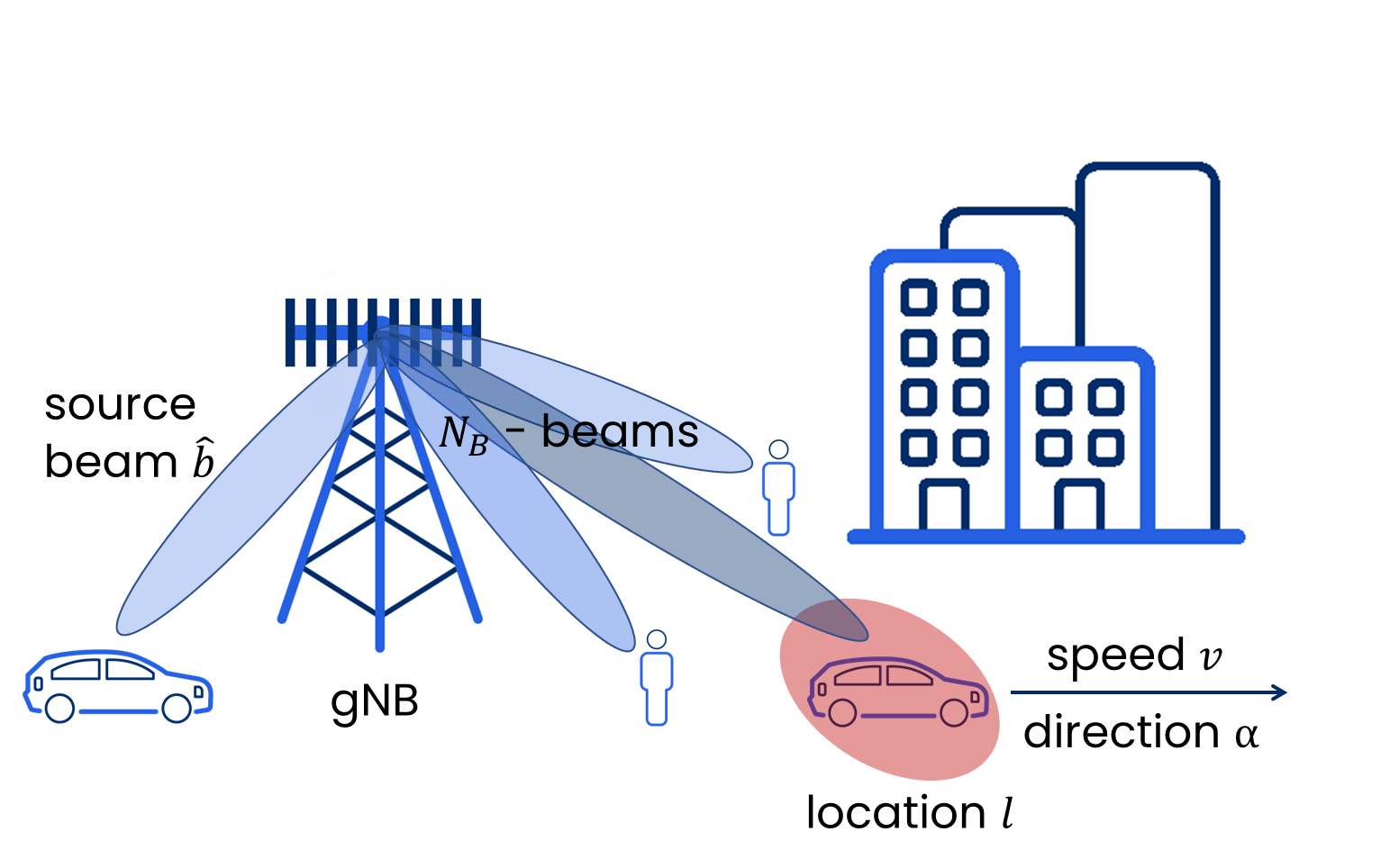}}
\caption{System model}
\label{fig:system_model}
\end{figure}

In this paper, we are considering a BM procedure in a single cell of a 5G network, as depicted in Fig.~\ref{fig:system_model}. In the cell, there is a single gNB equipped with an antenna array that supports~$N_{\mathrm{B}}$ beams. The SSB signals related to those beams are transmitted every $T_{\mathrm{B}}$~ms. During each SSB burst, UE being in location~$l$ can compute RSRP related to every beam $b$ denoted as $P_r(b,l)$. After each SSB burst, a decision can be made to switch UE between beams, indicated by the index of a target beam $b$. Following the definitions in~\cite{Abinader2021} we assume that \emph{radio link failure} is the situation when the used source beam $\hat{b}$ has RSRP~$P_r(\hat{b},l)$ smaller by at least $\delta_\mathrm{th}$ than the strongest beam:
\begin{equation}\label{eq:best_beam}
    P_r(\hat{b},l) < \max_b\left(P_r(b,l)\right) - \delta_\mathrm{th}.    
\end{equation}

The considered cell implements the baseline BM procedure proper for the 5G network as defined in~\cite{Abinader2021}. The beam is switched when RSRP related to the best target beam exceeds the RSRP of the source beam by at least $\delta_\mathrm{ho}$ (defined margin): 
\begin{equation} \label{eq:lte_ho_1}
    P_r(\hat{b},l) < \max_b\left(P_r(b,l)\right) - \delta_\mathrm{ho}.
\end{equation}
While the RSRP measurement is available after the SSB burst, the baseline BM procedure suffers a delay of $T_{\mathrm{B}}$~ms.

\section{Proposed BM implementation in ORAN} \label{sec:oran_architecture}
\begin{figure}[htbp]
\centerline{\includegraphics[height=4.5cm]{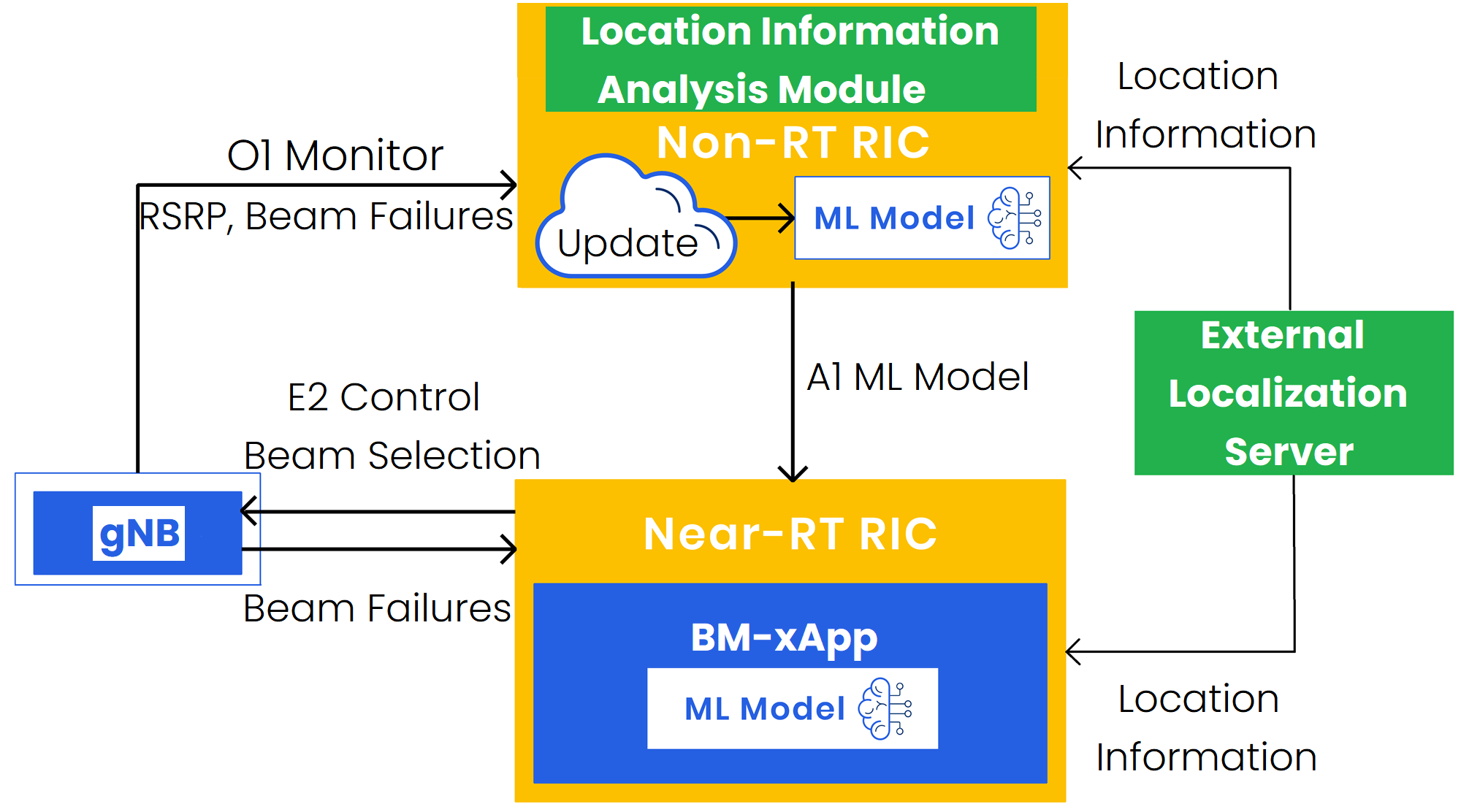}}
\caption{Beam Management using the O-RAN architecture}
\label{fig:bmm_oran}
\end{figure}

The aim of O-RAN is to standardize interfaces between RAN components, allowing MNOs to deploy a network consisting of components provided by various vendors. Moreover, the architecture is extended by the RAN Intelligent Controller (RIC). RIC is a unit that is responsible for the reconfiguration of RAN, based on the data collected from RAN components, and processed with the use of Machine Learning (ML) techniques. 
It is assumed that the cell considered in this paper (see Sec.~\ref{sec:system_model}) is compliant with the O-RAN architecture~\cite{dryjanski2021toward} depicted in Fig.~\ref{fig:bmm_oran}. The BM functionality is deployed within RIC. RIC is split into two parts: Non-RT RIC, which operates in the time-frame of above 1~s, and Near-RT RIC, which operates in near real-time, i.e., between 10~ms and 1~s~\cite{dryjanski2021toward}. The O-RAN specification defines the roles of both Non-RT RIC and Near-RT RIC in the case of BM~\cite{oran2022}. Non-RT RIC is proper for, e.g., data capture and aggregation, training of ML models, and extensive computations. Depending on the implementation, Non-RT RIC modules can be either rApps or built-in vendor-dependent entities. On the other hand, Near-RT RIC hosts the so-called BM-xApp that performs near real-time actions in the network, i.e., reselects beams for UEs. The communication between Non-RT RIC and Near-RT RIC is performed through the A1 interface. The A1 interface provides mechanisms to, e.g., send ML models obtained in Non-RT RIC to Near-RT RIC to make inference. BM-xApp controls beam reselections through the E2 interface attached to gNB. The E2 interface can be used also to obtain feedback information about, e.g., \emph{radio link failures}. On the other hand, the information from gNB can be obtained in Non-RT RIC via the O1 interface. It can be, e.g., RSRP reported by the users, and \emph{radio link failure} statistics. The observed increased number of {\emph{radio link failures}} can trigger an update of the ML model in Non-RT RIC. At the same time, BM-xApp in Near-RT RIC can trigger mechanisms to prevent Quality of Service (QoS) degradation, e.g., temporal utilization of the baseline switching algorithm until the ML model inference is completed in the changed radio environment. In addition, both Non-RT RIC and Near-RT RIC can use external services like a Localization Server. According to the O-RAN specification~\cite{oran2022}, the Localization Server provides information about user location captured from the external application layer to both Non-RT RIC and Near-RT RIC. We expect Non-RT RIC to contain a module dedicated to long-term analysis of user location information in order to extract, e.g., UE mobility patterns. 

\section{REM-Based Beam Management} \label{sec:rem-bm}

Following the O-RAN architecture, the proposed REM-based BM is split between Non-RT RIC and Near-RT RIC. The first part of this section describes the deployment of REM-based BM related to Non-RT RIC, while the latter one describes the Near-RT RIC part of REM-based BM.

\subsection{Non-RT RIC}

\begin{figure}[htbp]
\centerline{\includegraphics[height=3.5cm]{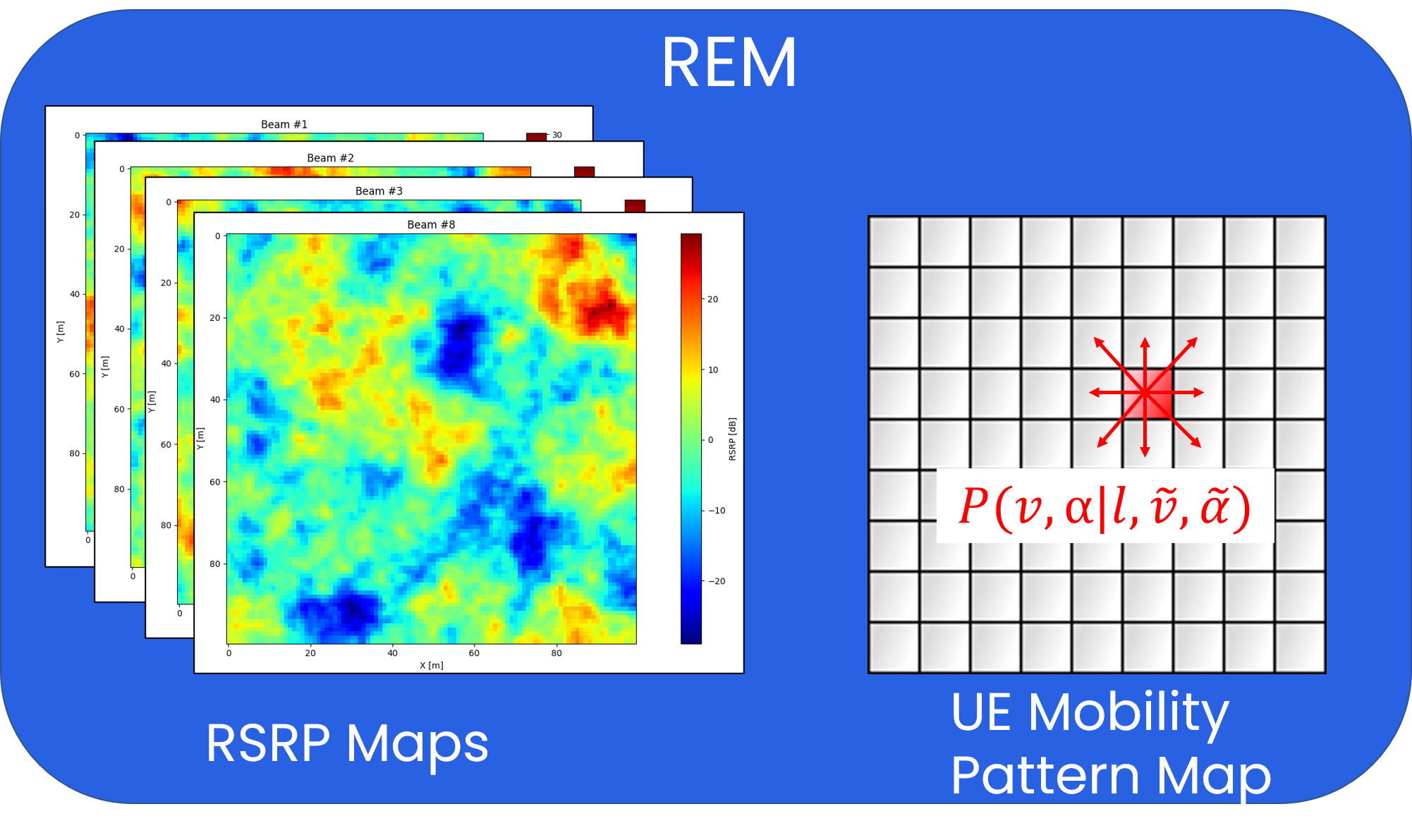}}
\caption{Data stored in REM to be used for BM.}
\label{fig:rem}
\end{figure}

Non-RT RIC operates in a time frame of above~1~s, thus it is appropriate for the implementation of REM-based BM functionality that requires extensive computations, and large memory. The first element of the proposed algorithm to be implemented in Non-RT RIC is REM. REM stores RSRP values for each location and beam as reported by UEs creating maps, as depicted in Fig.~\ref{fig:rem}. The input RSRP data is provided to Non-RT RIC through the O1 interface and is associated with the UE location provided by the external Localization Server. Such a location-RSRP pair is used to update REM. REM is built on the basis of a square grid with a single tile having a size equal to $g$, e.g., $10 m$. The reported RSRP values are averaged within each REM tile for a given beam. Besides the RSRP map, REM contains a map of UE mobility pattern over the cell area. 
The UE mobility pattern is given as the conditional probability~$\mathcal{P}(v,a|l,\tilde{v}, \tilde{\alpha})$ of moving with speed~$v$ in direction~$\alpha$, while being in location~$l$, and moving currently with the speed of $\tilde{v}$ in direction $\tilde{\alpha}$.
While in many cases this model allows us to predict UEs future velocity with high reliability, e.g., for cars moving on a motorway with high probability, it can be used to model more difficult traffic situations as well, e.g., UEs on a junction turning left and right with 40\% and 60\% probability, respectively.
This mobility map is obtained through a long-term analysis of the location information (position, speed, direction) provided by an external Localization Server in a dedicated Non-RT RIC module. These can be used to empirically estimate probabilities for the UEs mobility map. To reduce the amount of data being stored in REM, some quantization is necessary, e.g., position rounded to fit into one of REM's square tiles, direction split into 8 possible angles of $45$~deg. resolution. Under some specific scenarios, e.g., a small cell dedicated for users inside high-speed vehicles traveling along the motorway, or while initializing the UE Mobility Pattern Map for a given tile, simple estimation, i.e.,  $v = \hat{v}$ and $\alpha = \hat{\alpha}$,  can be sufficient.

It can be seen that the definition of a UE mobility pattern specified by probability $\mathcal{P}(v,a|l,\tilde{v}, \tilde{\alpha})$ follows the Markov property, i.e., the next UE speed $v$ and direction $\alpha$ depends only on the present speed $\tilde{v}$, direction $\tilde{\alpha}$, and location $l$. We assume that these do not depend on the previous state. From this perspective, beam selection in REM-based BM can be formulated as the Markov Decision Process (MDP), where a UE moves over the cell area, and sequential decisions are made by BS on its association with a certain beam. Based on the knowledge from REM these decisions can be assessed in terms of expected RSRP values at a given location. The algorithm iteratively learns, off-line the optimal beam assignment, in terms of a given goal function. The defined MDP consists of the following components:
\begin{itemize}
    \item \textbf{Agent} is a UE that moves over the cell area following its mobility pattern $\mathcal{P}(v,\alpha|l,\tilde{v}, \tilde{\alpha})$. In each state $s$, the Agent takes one of the available actions $a$ on the basis of policy~$\pi(s, a)$
    \item \textbf{Environment} is the considered cell. The cell (environment) is modeled through REM, namely the RSRPs related to particular SSBs at a given UE location $P_r(b,l)$, and UE's mobility pattern at a given location $\mathcal{P}(v,a|l,\tilde{v}, \tilde{\alpha})$.
    \item \textbf{State} $s$ is defined as the current UE speed $\tilde{v}$, direction $\tilde{\alpha}$, location $l$, and source beam $\hat{b}$.
    \item \textbf{Action} $a$ is the selection of one of the $N_{\mathrm{B}}$ target beams
    \item \textbf{Policy} $\pi(s,a)$ is the probability distribution of taking action $a$ in state $s$.
    \item \textbf{Reward} is defined so as to enable MNO balancing between the minimization of the number of beam reselections, and using the beams of the highest RRSP. Its formal definition is:
    \begin{equation} \label{eq:reward}
        r(s,a) = \beta \cdot f_{\mathrm{BR}}(s,a) + (1-\beta) \cdot f_{\mathrm{RSRP}}(s,a),
    \end{equation}
    where $\beta \in [0,1]$ is the design parameter to balance between the minimization of beam reselections, and RSRP maximization goals, respectively. $f_{\mathrm{BR}}(s,a)$ is defined as:
    \begin{equation} \label{eq:min_ho}
        f_{\mathrm{BR}}(s,a) = \begin{cases}
        -1000 & \text{if  \emph{\eqref{eq:best_beam}} is true} \\
        -1 & \text{if \hspace{0.1cm}} a \neq \hat{b} \\
        0 & \text{otherwise}, 
        \end{cases} 
    \end{equation}
    and $f_{\mathrm{RSRP}}(s,a)$ is given by:
    \begin{equation} \label{eq:max_rsrp}
        f_{\mathrm{RSRP}}(s,a) = \begin{cases}
        0 & \text{if \hspace{0.1cm}} a = \arg  \max_b\left(P_r(b,l)\right) \\
        -1000 & \text{otherwise}.
        \end{cases} 
    \end{equation}
    In the case of $f_{\mathrm{BR}}(s,a)$, a high negative reward of $-1000$ is associated when an action causes a \emph{radio link failure}~\eqref{eq:best_beam}. A small negative reward of~$-1$ is given to the agent for switching the source beam~$\hat{b}$ to another beam, and a reward equal to~$0$ is given when the beam is not switched. This function is maximized if the number of beam switches is minimized while not causing radio link failures.  In the case of $f_{\mathrm{RSRP}}(s, a)$, reward equal to~$0$ is given to the agent for selecting the highest-RSRP  beam. Other actions are associated with large negative rewards of~$-1000$. The function is maximized if always the highest RSRP beam is selected.
    One should notice that $f_{\mathrm{BR}}(s,a)$, and $f_{\mathrm{RSRP}}(s,a)$ are only the representative examples of goal functions. The proposed solution can work with flexibly defined reward functions that will suit MNOs' specific needs.
\end{itemize}

By storing information about the environment dynamics (UE mobility patterns), and information about the rewards in states (to be computed from RSRP maps) in REM, we can obtain the optimal solution for MDP offline in Non-RT RIC using the so-called Policy Iteration (PI) algorithm~\cite{sutton2018reinforcement}. The aim of the PI algorithm is to obtain the optimal policy, i.e., the policy that maximizes the so-called expected discounted reward. The PI algorithm starts from an arbitrary policy and consists of two alternating steps, namely \emph{Policy Evaluation}, and \emph{Policy Improvement}. The \emph{Policy Evaluation} step aims at estimating the so-called \emph{value function}, which stands for the expected return while being in state $s$ and following policy $\pi(s,a)$ thereafter. After the \emph{Policy Evaluation} is finished, \emph{Policy Improvement} begins. It modifies the policy toward the optimal one. The algorithm is finished when the policy is not changing anymore, i.e., it converged to the optimum. 
\subsection{Near-RT RIC} \label{subsec:nearrt_ric}

Offline-computed optimal policies to be followed in each state $\pi^*(s, a)$ are transferred from Non-RT RIC to BM-xApp that is located in the near RT RIC, through a dedicated A1 ML interface (See Fig.~\ref{fig:bmm_oran}). The aim of BM-xApp is to enforce those policies, i.e., to indicate gNB through the E2 interface, that a particular UE should be switched to another target beam. BM-xApp determines state $s$ of the UE on the basis of its current location $l$, speed $v$, and direction $\alpha$ obtained from the external Localization Server. The information in REM, and consequently the policies, are quantized in space, i.e., following the square grid of size $g$. Thus the UE location obtained from the external Localization Server is rounded to fit the closest (in terms of Euclidean distance) point in REM constituting input location $l$. BM-xApp also provides a mechanism to prevent QoS degradation. This might be the case when, e.g., data in REM becomes outdated, or some temporal high RSRP variation occurs. If a \emph{radio link failure} is detected, BM-xApp arbitrary indicates through the E2 interface to gNB to switch this user to the new target beam providing the highest RSRP.

\section{Simulation Scenario} \label{sec:scenario}

In this paper, we are considering BM in a single small cell covering an area of $500 \times 500$ m. A single gNB is deployed in the middle of the left edge, i.e., of x-y coordinates $(0,250)$~m. The gNB is equipped with an $8 \times 8$ rectangular antenna array that is based on the \emph{Micro Urban} scheme described in the 5G specification~\cite{3gpp2022b}. The gNB supports $N_{\mathrm{B}}=16$ orthogonal beams. SSB burst period $T_{\mathrm{B}}$ equals $20$~ms, which is a typical value reported in the literature~\cite{barati2020energy, Na2019}. The antenna array is installed at the height of 10~m and transmits a signal with the TX power distributed to all beamforming antennas of $10$~dBm/MHz. The radiation pattern is computed following the formulas from~\cite{3gpp2022b}. The gNB operates with a bandwidth of $100$~MHz at the center frequency of $26$~GHz. Due to the high path loss, this band is suitable for such a small area cell. To model the impact of the radio environment on RSRP related to each beam, e.g., the presence of some obstacles that attenuate the signal, we have used a spatially correlated shadowing model independent for each beam~\cite{Zhang2012}. We have set the correlation distance to $10$~m. In order to ensure identical radio conditions for the evaluated algorithms, the same random seed is used to generate correlated shadowing before each simulation. The radio channel coefficients are affected by fast fading that is independent for each realization and follows the Rayleigh distribution. BM is especially challenging for high-speed users. Thus, we are considering the urban road scenario, where UEs are moving with the speed of $v=25$~m/s. The road users can only move along the defined lanes, so in this scenario, the direction can be either $\alpha=0$~(downward) or $\alpha=180$~(upward). We assume that high-speed users utilize cm-level accuracy Real Time Kinematics localization, which is supported in mobile networks since LTE~\cite{hoffmann2020}. We are considering $300$~UEs going through the cell downward. We set the margin for \emph{radio link failure} to $\delta_\mathrm{th} = 8$~dB~\cite{Suarez2017}. The simulation parameters are summarized in Table~\ref{tab:simulation_parameters}. 

\begin{table}[htbp]
\caption{Simulation Parameters}
\begin{center}
\begin{tabular}{|c|c|}
\hline
\textbf{Parameter} & Value \\
\hline
cell size & $500 \times 500$~m \\
\hline
gNB placement (x,y) & $(0,250)$~m \\
\hline
height of antenna array & $10$~m \\
\hline
antenna array model & $8\times 8$ \emph{Micro Urban}~\cite{3gpp2022b} \\
\hline
number of beams $N_{\mathrm{B}}$ & 8 \\
\hline
SSB-burst period $T_{\mathrm{B}}$ & $20$~ms \\
\hline
TX power & $10$~dBm/MHz \\
\hline
center frequency & $26$~GHz \\
\hline
correlation distance for shadowing & $10$~m \\
\hline
number of users & $300$ \\
\hline
user speed $v$ & $25$~m/s \\
\hline
user direction $\alpha$ & $\{0,180\}$\\
\hline
margin for \emph{radio link failure} $\delta_\mathrm{th}$ & $8$ dB \\
\hline
\end{tabular}
\label{tab:simulation_parameters}
\end{center}
\end{table}

\section{Simulation Results} \label{sec:results}

The proposed REM-based BM is evaluated within the scenario described in Sec.~\ref{sec:scenario}. At first, we populated the REM with RSRP data, so that it contains a statistically correct estimation of the mean RSRP value, over all available reports, for each beam within every square tile of size $g=2$~m. Next, we created a UE mobility pattern map, based on the knowledge about the possible UE speed $v$ and directions $\alpha$. Having the necessary data collected, we ran the PI algorithm to obtain policies. We obtained two sets of policies: the first, are aimed at the minimization of beam reselections, while avoiding \emph{radio link failures}~\eqref{eq:best_beam} (BR-MIN). Second, aimed at RSRP maximization (RSRP-MAX). The corresponding $\beta$ values are $1$, and $0$ for BR-MIN, and RSRP-MAX, respectively. We compared the proposed solutions against the standard BM procedures in 5G (denoted as \emph{Baseline}) with $\delta_\mathrm{ho}$ values  of $\{3,5,7 \}$~dB.
\begin{figure}[htbp]
\centerline{\includegraphics[height=4.0cm]{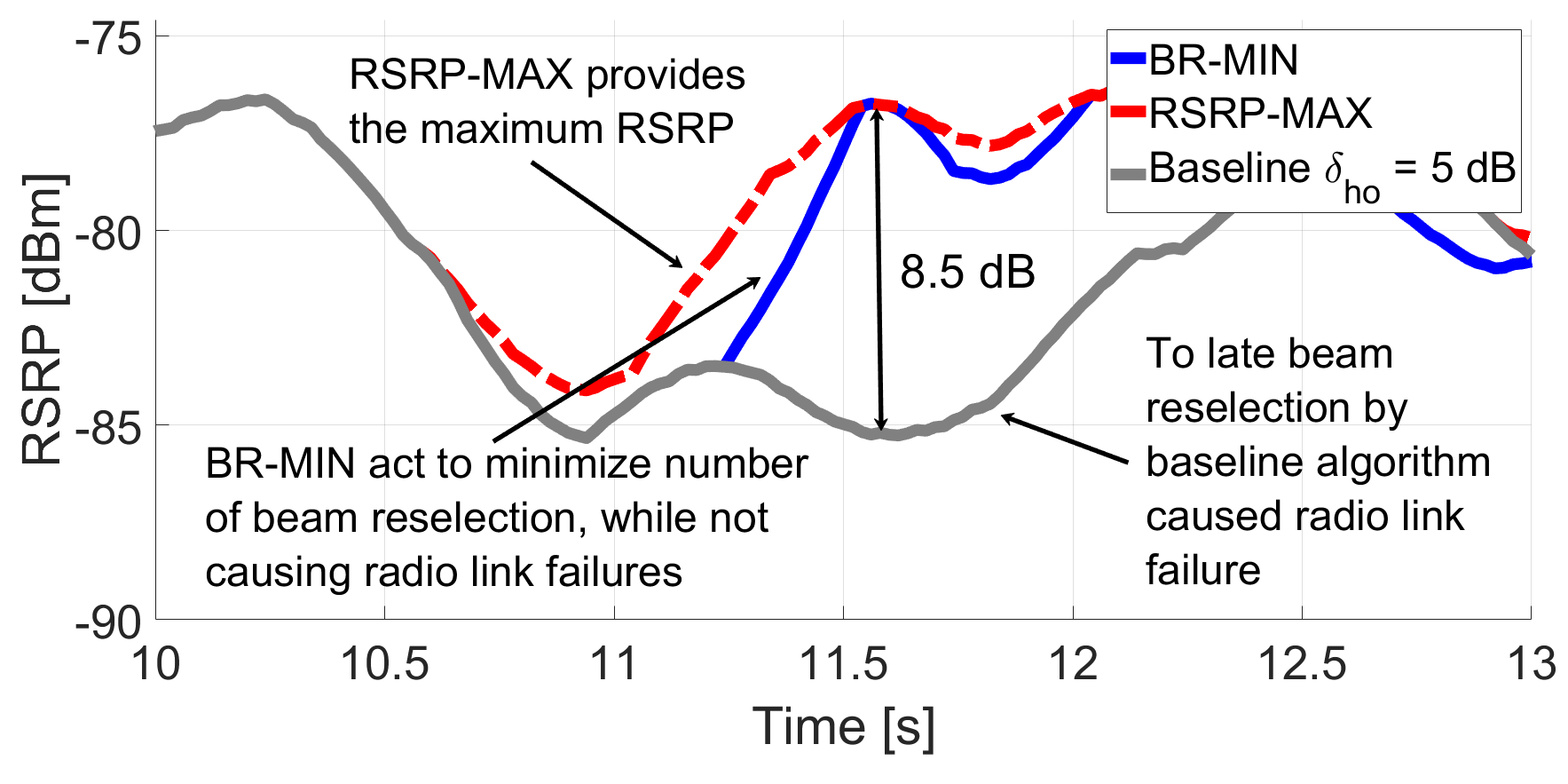}}
\caption{Illustration of RSRP-MAX, BR-MIN, and \emph{Baseline} ($\delta_\mathrm{ho}=5$~dB) algorithms. The results are obtained for a single user, and a 15-point moving average is applied to smooth the plot. }
\label{fig:demo}
\end{figure}
In Fig.~\ref{fig:demo} there is an illustration of RSRP-MAX, BR-MIN, and \emph{Baseline} ($\delta_\mathrm{ho}=5$~dB) algorithms in the form of an RSRP vs. time plot for a single user. To smooth out the results, a 15-point moving average is applied. The RSRP-MAX algorithm provides the user with the highest RSRP based on information in REM. On the opposite, the \emph{Baseline} algorithm to prevent the ping-pong effect performs beam reselection when a better target beam, i.e., of RSRP above the current by at least $5$~dB, is detected. It can be seen that for a high-speed user it causes \emph{radio link failures}. It is due to its reactive character, i.e., beam reselection can be triggered only after a new RSRP measurement arrives, i.e., after $T_{\mathrm{B}}$~ms. BR-MIN proactively minimizes the number of beam reselections. Due to the historical knowledge from REM, it is trained to perform beam reselection before the RSRP of the source beam significantly drops (and is reported), causing \emph{radio link failure}.

\begin{figure}[htbp]
\centerline{\includegraphics[height=4.2cm]{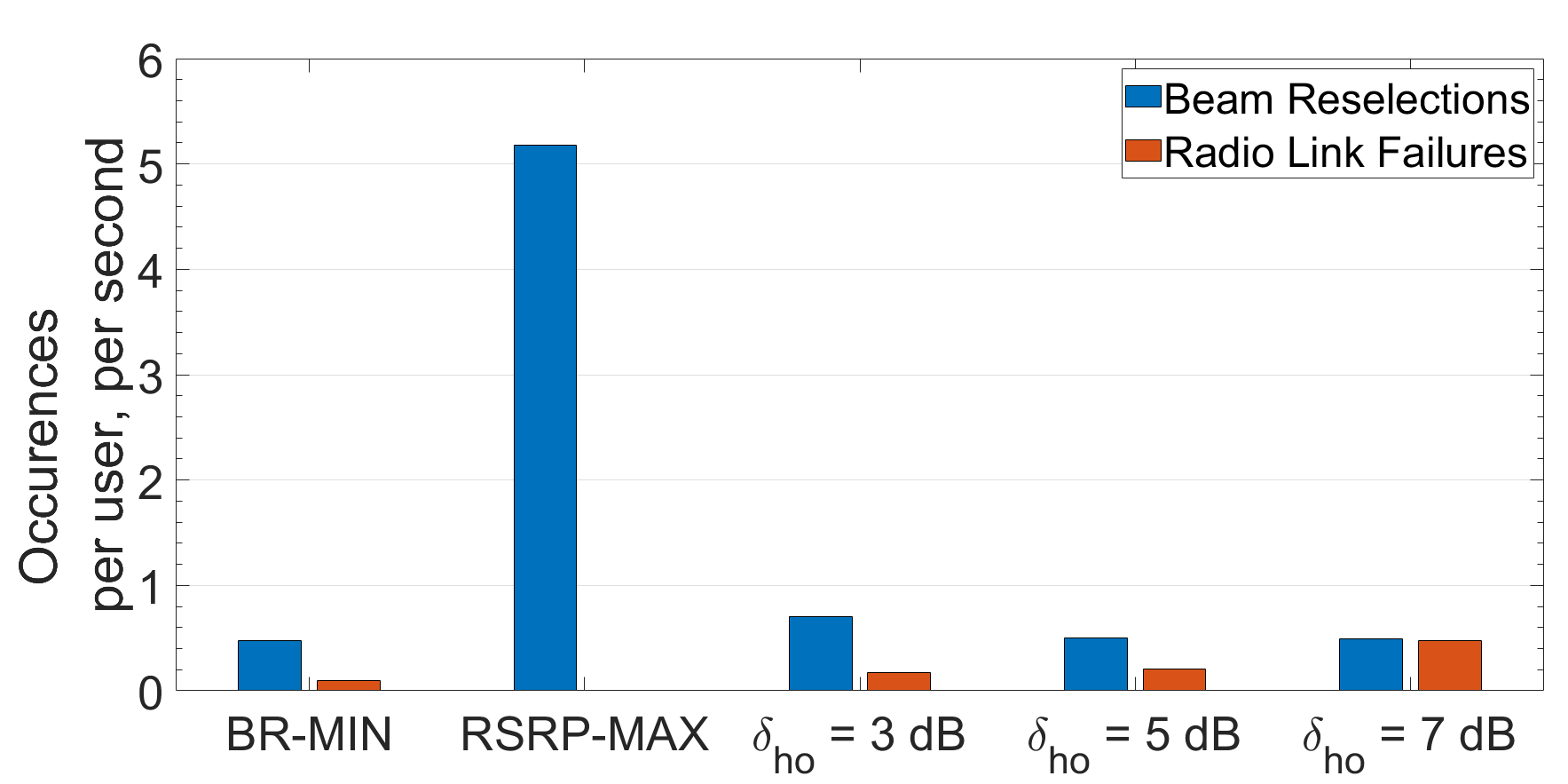}}
\caption{Beam reselections and \emph{radio link failures}. Averaged per user, per second. }
\label{fig:beam_reselections}
\end{figure}

To capture statistically significant KPIs related to the performance of the proposed REM-based BM, it is compared with the \emph{Baseline} approach exploiting different values of~$\delta_\mathrm{ho}$. A 15~s long simulation has been performed for all the considered 300 UEs. The first comparison is aimed at the number of beam reselections, and the number of \emph{radio link failures}. The results are presented in Fig.~\ref{fig:beam_reselections}. The statistics are averaged over time and the number of UEs, i.e., the number of occurrences per user per second.  
It can be seen that when BM-xApp acts according to the RSRP-MAX the number of beam reselections is the highest. It is reasonable, as the algorithm constantly follows the beam providing the highest RSRP. BM-xApp following the BR-MIN policy is characterized by about~29\% reduction of the number of beam reselections compared to the \emph{Baseline} algorithm with~$\delta_{\mathrm{ho}} = 3$ dB. For the remaining values of~$\delta_{\mathrm{ho}}$, the \emph{Baseline} algorithm has similar beam reselection performance, i.e., about 0.5 reselections per user per beam. However, the \emph{Baseline} algorithm has much poorer statistics of \emph{radio link failures}, i.e., $1.81$, $2.08$, and $4.91$ times worse for $\delta_{\mathrm{ho}}$ = $3$, $5$, and $7$~dB, respectively. It is because RSRP related to a particular beam can decay fast, especially when UE moves at a high speed. BM-xApp utilizes policies obtained from REM, thus it can predict the time of beam reselection to avoid QoS deterioration. However, due to fast fading, sometimes temporal channel variations can cause a \emph{radio link failure}, even when the BR-MIN policy is used. 

\begin{figure}[htbp]
\centerline{\includegraphics[height=4.5cm]{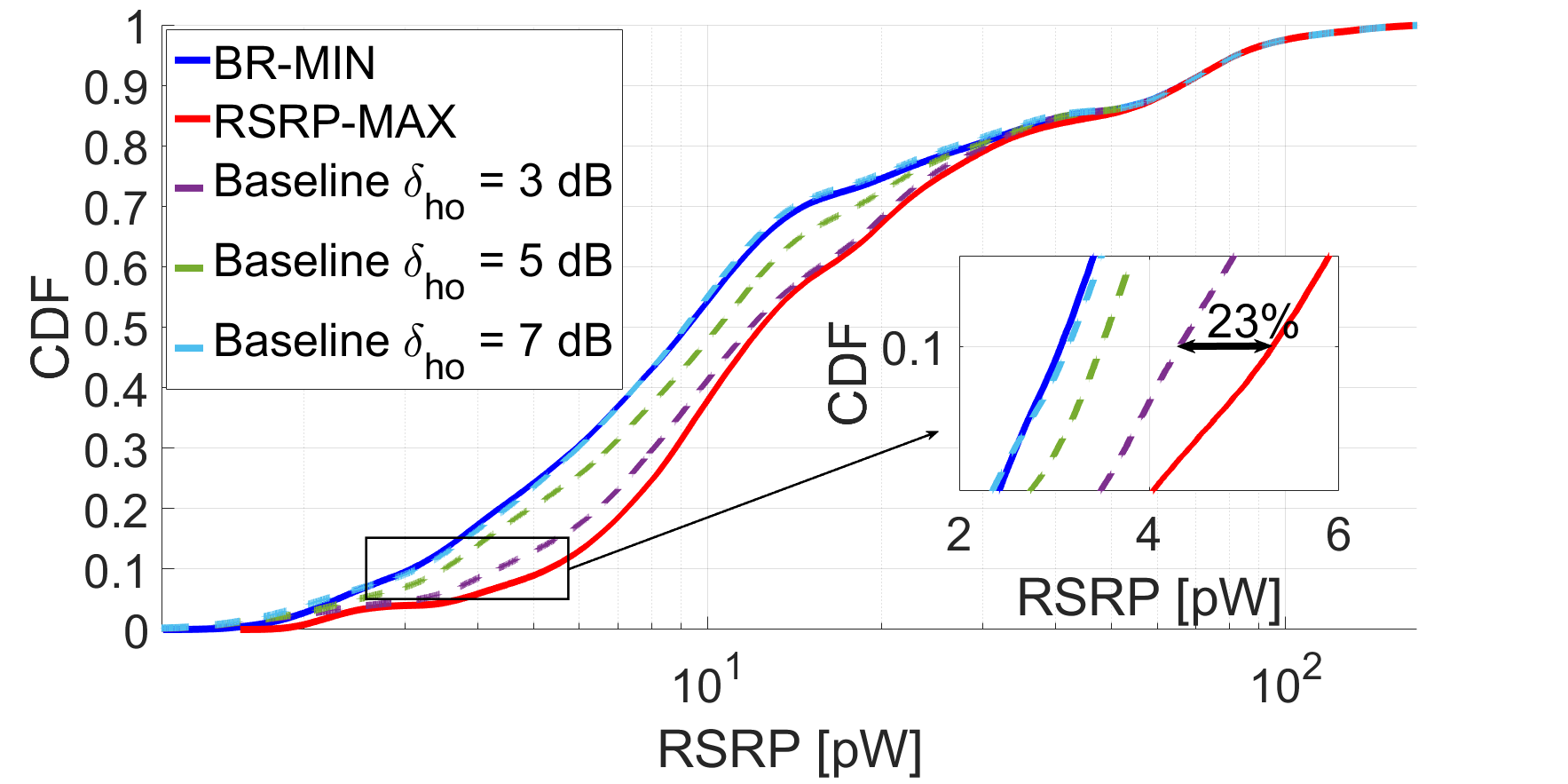}}
\caption{Comparison of RSRP distributions related to the BM xApp, and the \emph{Baseline} algorithms. }
\label{fig:rsrp}
\end{figure}

Another KPI to compare between BM-xApp and the \emph{Baseline} algorithm is RSRP distribution. The Cumulative Distribution Function (CDF) of RSRPs observed by all of the UEs during the 15~s long simulation is depicted in Fig.~\ref{fig:rsrp}. It can be seen that BM-xApp following the RSRP-MAX policy always chooses the beam with the highest RSRP. The highest gains can be observed for the so-called cell-edge users, i.e, the 10th percentile of the RSRP distribution. The RSRP-MAX algorithm improves RSRP of the cell-edge users by about 23\% compared to the \emph{Baseline} algorithm (for $\delta_{\mathrm{ho}}= 3$~dB). On the other hand, when BM-xApp follows the BR-MIN policy, RSRP significantly drops. Although the RSRP distribution in the case of BM-xApp following the BR-MIN policy is similar to the \emph{Baseline} algorithm for $\delta_{ho}=7$~dB, it is important that, unlike with the \emph{Baseline} algorithm, decisions made by BM-xApp rarely cause \emph{radio link failures}.

\section{Conclusions} \label{sec:conclusions}

In this paper, we have proposed BM based on the information stored in REM. It is close to practical implementation because of using the O-RAN architecture. In the paper, we have clearly defined the role of Non-RT RIC, and Near-RT RIC, together with interfaces for their communications, in the context of the proposed BM. 
The BM xApp was tested for two policies, i.e., BR-MIN, and RSRP-MAX, against the baseline algorithm. Simulation studies have shown that following the BR-MIN policy can minimize the number of beam reselections, while preventing QoS degradation caused by rapid drops of RSRP, e.g., for high-speed UEs. While the \emph{baseline} algorithm can obtain similar number of beams reselections it is penalized by $2.18$ times more \emph{radio link failure} events. On the other hand, the proposed BM following the RSRP-MAX policy provides users with maximum RSRP values. Most importantly, the proposed solution is flexible, i.e., goal functions can be redefined according to the MNOs' needs.

\bibliography{bibliography} 
\bibliographystyle{IEEEtran}

\end{document}